\documentclass{article}
\usepackage{spconf,amsmath,graphicx}

\newlength\savedwidth

\usepackage{multirow}
\usepackage{amsmath}
\usepackage{amsfonts}
\usepackage{colortbl}
\usepackage{cite}
\usepackage{arydshln}
\usepackage{url}
\usepackage{algorithmic}
\usepackage{algorithm}

\newcommand{\ten}[1]{ \boldsymbol{\mathcal #1}}
\newcommand{\bbR}[1]{\mathbb{R}^{#1}}
 
\newcommand{\bm}[1]{{\mbox{\boldmath $#1$}}}

\makeatletter
\def\bstctlcite{\@ifnextchar[{\@bstctlcite}{\@bstctlcite[@auxout]}}
\def\@bstctlcite[#1]#2{\@bsphack
\@for\@citeb:=#2\do{%
\edef\@citeb{\expandafter\@firstofone\@citeb}%
\if@filesw\immediate\write\csname #1\endcsname{\string\citation{\@citeb}}\fi}%
\@esphack}
\makeatother

\usepackage{color}

\title{Low-rank constrained multichannel signal denoising\\considering channel-dependent sensitivity inspired by self-supervised learning for optical fiber sensing}

\name{\begin{tabular}{c}Noriyuki Tonami$^{1}$ Wataru Kohno$^{2}$, Sakiko Mishima$^{1}$,\\
Yumi Arai$^{1}$, Reishi Kondo$^{1}$, Tomoyuki Hino$^{1}$\end{tabular}}
\address{$^1$NEC Corporation, Japan, $^2$NEC Laboratories America, Inc., America}

\begin{document}
\ninept
\maketitle
\begin{abstract}
	
Optical fiber sensing is a technology wherein audio, vibrations, and temperature are detected using an optical fiber; especially the audio/vibrations-aware sensing is called distributed acoustic sensing (DAS).
In DAS, observed data, which is comprised of multichannel data, has suffered from severe noise levels because of the optical noise or the installation methods.
In conventional methods for denoising DAS data, signal-processing- or deep-neural-network (DNN)-based models have been studied.
The signal-processing-based methods have the interpretability, i.e., non-black box.
The DNN-based methods are good at flexibility designing network architectures and objective functions, that is, priors.
However, there is no balance between the interpretability and the flexibility of priors in the DAS studies.
The DNN-based methods also require a large amount of training data in general.
To address the problems, we propose a DNN-structure signal-processing-based denoising method in this paper.
As the priors of DAS, we employ spatial knowledge; low rank and channel-dependent sensitivity using the DNN-based structure.
The result of fiber-acoustic sensing shows that the proposed method outperforms the conventional methods and the robustness to the number of the spatial ranks.
Moreover, the optimized parameters of the proposed method indicate the relationship with the channel sensitivity; the interpretability.

\end{abstract}

\vspace{0pt}
\section{Introduction}
\label{sec:intro}
\vspace{0pt}

Distributed Acoustic Sensing (DAS), which is also known as coherent optical time-domain reflectometry (OTDR) or $\phi$-OTDR, is one of optical fiber sensing \cite{IP_OFC2021,Ip_JOCN2022}.
In DAS, sounds and/or vibrations are captured by using optical fibers.
As the other feature of DAS, large-scale multichannel data can be obtained along the optical fiber; the data includes much spatial information.
The variations of the observation have brought about various types of applications, such as whale call detection \cite{Bouffaut_Frontiers2022}, structural health monitoring \cite{Peter_JCSHM2021}, seismic activity monitoring \cite{Parker_FirstBreak2014}, border monitoring \cite{Owen_EISIC2012}, large-scale sensing over 171km \cite{Waagaard_OSAC2021} and pole localization \cite{Lu_ICASSP2021}.

Denoising observed signals by DAS has been in the spotlight \cite{Bekara_GEOPHYSICS2007,Chen_GEOPHYSICS2014,Zhao_GJI2021,Feng_TGARS2022,Yang_GEOPHYSICS2023}.
The denoising brings downstream tasks, e.g., classifying events, to the improvement of the performance.
In DAS, the signal-to-noise ratio (SNR) for each channel is different because of the factors in the optical domain \cite{Zhou_JLT2013} and installation methods such as aerial or buried.
Bekara {\it et al.} have applied low-rank approximation to denoise DAS data.
DAS data is comprised of many sensing points.
Low-rank approximation in the spatial dimensions in DAS is helpful in mitigating random noise, e.g., optical shot noise.
The above signal-processing-based methods are advantageous for understanding the behavior of parameters, i.e., white box.
On the other hand, the performance might be limited because of the flexibility of the model parameters.
Recently, deep neural network (DNN) has boosted the denoising performance for DAS \cite{Zhao_GJI2021,Feng_TGARS2022,Yang_GEOPHYSICS2023}, which outperform the signal--processing--based methods. 
In the DNN methods, various types of prior information can be included in the DNN structures and objective functions.
This is because DNN models require only differentiable functions as the structures and objective functions, which leads to the flexibility of the prior information for the models.
The disadvantage is that DNN-based methods require large training data.
To tackle this problem, deep image prior (DIP) \cite{Lempitsky_CVPR2018} has been studied for the denoising.
In DIP, the iterative online denoising is performed using only one noisy sample with DNN structures as prior information.
However, the DNN-based methods, including DIP, still suffer from the black box.

To tackle these problems, we propose a DNN-structure signal-processing-based denoising where none of the training data are required and which is a non-black box.
In the proposed denoising method, only one noisy sample is required, i.e., no machine learning.
Moreover, the priors of DAS are easily leveraged because of the flexibility of the proposed DNN-structure-based signal processing.
In this work, as the priors, we invoke the spatial characteristics of DAS; the low rank and channel-dependent sensitivity.

\begin{figure*}[t!]
  \centering
  \includegraphics[width=0.93\textwidth]{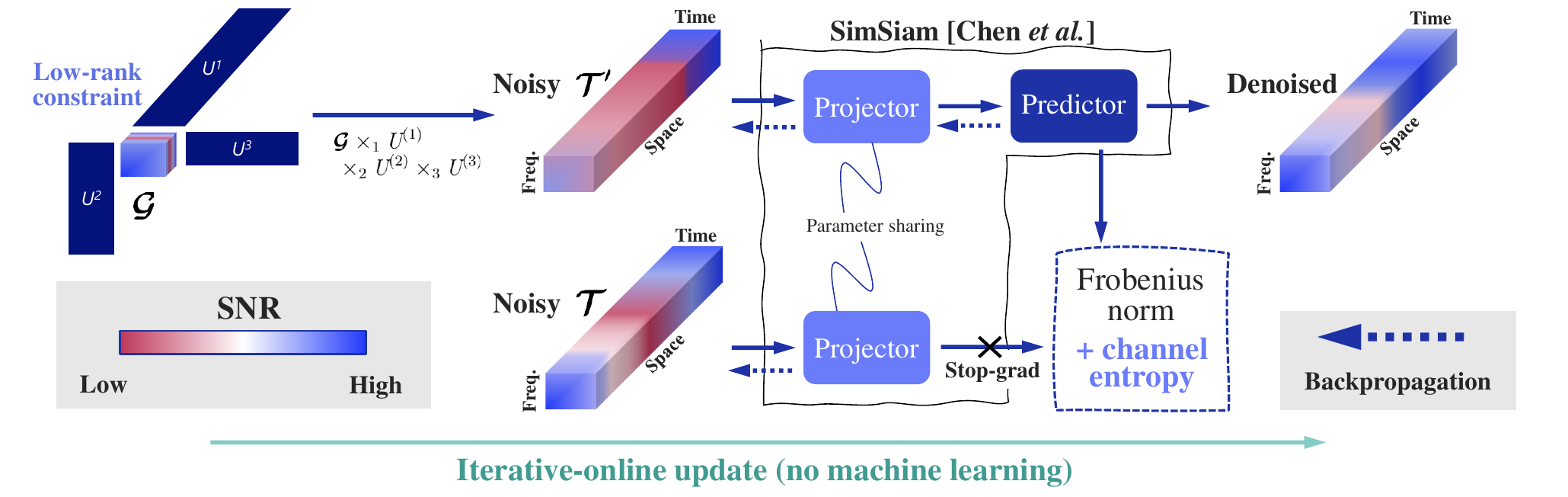}
  \vspace{-5pt}
  \caption{ Graphical repsesentation of proposed method }
  \label{fig:prop_image}
  \vspace{0pt}
\end{figure*}

\vspace{0pt}
\section{DAS of optical fiber sensing}
\vspace{0pt}

\subsection{Principle of DAS}
\vspace{0pt}
In DAS, a phase change of Rayleigh backscattering light wave is captured along an arbitrary location of a fiber cable.
The backscattering occurred by a pulsive laser input into the cable.
The phase change is proportional to the dynamic strain of the cable \cite{Zhong_PTL2019}, that is, sounds or vibrations that propagate the optical fiber.
We then obtain a spatiotemporal signal by DAS as follows:
\begin{align}
  \vspace{0pt}
  \bm{\Delta \phi}_t &=  [ \Delta \phi_{t,0}, \dots, \Delta \phi_{t,c}, \dots, \Delta \phi_{t,C-1}] \in \mathbb{R}^{C},
  \label{eq:DAS_data}
  \vspace{0pt}
\end{align}

\noindent
where $\Delta \phi_{t,c}$ indicates the phase change, i.e., sounds and/or vibrations, at time $t$ and channel $c$, where the temporal interval depends on a period of launching the laser.
$C$ denotes the number of spatial channels, where its interval depends on the time of the flight of the laser in the cable.
Moreover, 3rd-order tensor $\ten{T} \in \bbR{I_c \times I_f \times I_t}$ can be obtained by an amplitude/power spectrogram for each channel with short-time Fourier transform (STFT).
$I_c$, $I_f$, and $I_t$ are respectively the numbers of channels, frequency bins, and time frames.

\begin{algorithm}[t]
	\caption{Proposed denoising} \label{alg:prop}
	\begin{algorithmic}[1]
		\STATE {\bf input:} Multichannel noisy signal $\ten{T} \in \bbR{I_c \times I_f \times I_f}$ and predefined rank  $(R_c,  R_f, R_t)$. $N$ is the number of iterations.
		\STATE {\bf initialize:} $\ten{G} \in \bbR{R_c \times R_f \times R_t}$, and $ \bm U^{(m)} \in \bbR{I_m \times R_m} \mid  {m= c,f,t } $ using Tucker decomposition of $\ten{T}$. 
	 $\beta_{c}$ and $\gamma_{c}$ are intitialized with zero and one, respectively.
	 The other parameters are randomly initialized.
		\FOR{$N$}
		\STATE $\ten{T}' \leftarrow \ten{G} \times_{1} {\bm U}^{(1)} \times_{2} {\bm U}^{(2)} \times_{3} {\bm U}^{(3)}$
		\STATE $\ten{Z}' \leftarrow \textsf{\textcolor{black}{Projector}}(\ten{T}', \bm \beta, \bm \gamma)  $
		\STATE $\ten{Z} \leftarrow \textsf{\textcolor{black}{Projector}}(\ten{T}, \bm \beta, \bm \gamma)  $ 
		\STATE $\ten{T}' \leftarrow \textsf{Predictor}(\ten{Z}' )  $
		\STATE $\ten{T} \leftarrow \textsf{Predictor}(\ten{Z} )  $ 
		\STATE $  \mathcal{L} \leftarrow \mathcal{L}_{\rm F}(\ten{T}, \ten{Z}', \ten{T}', \ten{Z}) + \mathcal{L}_{\rm ChEnt}(\bm \beta, \bm \gamma) $
		\STATE $ {\rm do}~\textsf{BackPropagation}$ 
		\STATE // Note: $\ten{Z}$ and $\ten{Z}'$ are not backpropagated.
		\ENDFOR
		\STATE {\bf output:} $(\ten{T} + \ten{T}')/2$;
	\end{algorithmic}
	\end{algorithm}

\vspace{0pt}
\subsection{Spatial characteristics of DAS data}
\vspace{0pt}
There are two characteristics of DAS data in terms of the spatial domain: the low rank and channel-dependent sensitivity.
The first characteristic is the low rank.
In DAS, a large amount of channels are obtained along an optical fiber.
A large-scale multichannel signal is produced when a long optical fiber is used for DAS.
In other words, the number of source signals is less than the number of channels, i.e., the low rank in the spatial domain.
The second characteristic is the channel-dependent sensitivity.
In DAS, the sensitivity of channels depends on the installation environments, e.g., aerial cables or subsea cables, and optical factors \cite{Zhou_JLT2013}.
In some channels, observed signals might be degraded because of the above reasons.

For considering the low rank in the spatial domain, matrix/tensor factorization is used \cite{Tucker_psy1996}.
As an example of the tensor factorization, Tucker decomposition is used \cite{Tucker_psy1996}.
In Tucker decomposition, given a 3rd-order tensor $\ten{Y} \in \bbR{I_1 \times I_2 \times I_3}$, $\ten{Y}$ is decomposed into a core tensor $\ten{G} \in \bbR{R_{1} \times R_{2} \times R_{3}}$ and three factors $\bm U^{(m)} \in \bbR{I_m \times R_m} \mid {m= 1,2,3 } $.
$R_{1}$, $R_{2}$, $R_{3}$ are respectively same with or less than $I_{1}$, $I_{2}$, $I_{3}$. 
Moreover, as for Tucker decomposition, the following special operation appears.
Given a tensor $\ten{X} \in \bbR{I_1 \times I_2 \times \dots \times I_n}$, a mode-$k$ multiplication between $\ten{X}$ and a matirix ${\bf A}$ $\in$ $\bbR{R \times I_{k}}$ is represented by $\ten{X} \times_{k} {\bf A} \in \bbR{I_1 \times \cdots \times I_{k-1} \times R \times I_{k+1} \times \cdots I_N} $.
``$\times_{k}$'' is the mode-$k$ multiplication.

\vspace{0pt}
\section{Proposed method}
\vspace{0pt}
To take advantage of both the signal processing and DNN; the interpretability and flexibility, the signal-processing-based denoising algorithm is implemented into the DNN-based structure.
For the signal-processing-based model, we utilize spectral subtraction \cite{berouti_icassp1979}, which brings about the white-box denoising method.
For the DNN-based structure, SimSiam \cite{Chen_CVPR2021} is utilized.
The combination of the signal-processing- and DNN-based methods enables us to perform denoising from only noisy signals with the interpretability and the flexible use of the priors of DAS.
Fig. \ref{fig:prop_image} shows the graphical representation of the proposed denoising method.
Algorithm \ref{alg:prop} is the algorithm of the proposed method represented by Fig. \ref{fig:prop_image}.

\vspace{0pt}
\subsection{Signal processing model}
\vspace{0pt}
Given a multichannel amplitude spectrogram $\ten{T} \in \bbR{I_c \times I_f \times I_t} \geq 0 $ of a noisy signal.
Assume that the background noise of the installation environments and/or the optical domain depends on only each channel, the noise can be subtracted from noisy signals $\ten{T} \in \bbR{I_c \times I_f \times I_t}$ as follows: 
\begin{align}
  \vspace{-5pt}
  {\sf Projector}(\ten{T}, \bm \beta, \bm \gamma) = \varphi \Big( \big(\ten{T}_{c,f,t} - \beta_{c}\big) / 
	\big(|\bm \gamma_{c} | + \epsilon\big) \Big) , \forall c, \forall f, \forall t\
  \label{eq:ch_subtraction}
  \vspace{-5pt}
\end{align}

\noindent $\varphi(\cdot)$ is an activation function, which is the hyperbolic tangent in the proposed method.
$\beta_c \in \mathbb{R}$ and $\gamma_c \in \mathbb{R}$ are respectively initialized with 0 and 1.
$\ten{T'}$ is also processed as shown in algorithm \ref{alg:prop} with Eq. \ref{eq:ch_subtraction}.
$\epsilon$ indicates the machine epsilon.
Eq. \ref{eq:ch_subtraction} is expected to extract the clean signal, which is similar to spectral subtraction \cite{berouti_icassp1979}.

\vspace{0pt}
\subsection{DNN-based structure incorporating priors of DAS}
\vspace{0pt}
Inspired by and combining the ideas of \cite{Chen_CVPR2021} and \cite{Bacca_AO2021}, we utilize the priors of DAS.
By utilizing the framework of SimSiam \cite{Chen_CVPR2021}, common features can be extracted from multiple inputs with avoiding the trivial solution in a self-supervised manner.

We mainly minimize the following function for incorporating one of the DAS priors; the low rank in the spatial domain.
\begin{align}
  \vspace{0pt}
  \mathcal{L}_{\rm F}(\ten{T}, \ten{Z}', \ten{T}', \ten{Z}) = (|| \ten{T} - \ten{Z}' ||_{\rm F}  +  || \ten{T}' - \ten{Z} ||_{\rm F})/2 \\
	s.t. ~~ \ten{T'} = \ten{G} \times_{1} {\bm U}^{(1)} \times_{2} {\bm U}^{(2)} \times_{3} {\bm U}^{(3)}
  \label{eq:Frobenius_norm}
  \vspace{0pt}
\end{align}

\noindent
$||\cdot||_{\rm F}$ denotes the Frobenius norm.
$\ten{G} \in \bbR{R_{1} \times R_{2} \times R_{3}}$ and $\bm U^{(m)} \in \bbR{I_m \times R_m} \mid {m= c,f,t } $ are a core tensor and three factors, respectively.
$R_{1}$, $R_{2}$, and $R_{3}$ is the predefined rank, respectively.
$\ten{Z}$ and $\ten{Z}'$ is the ouputs of {\sf Predictor($\cdot$)} \cite{Chen_CVPR2021}.
For {\sf Predictor($\cdot$)}, we use a single-layer point-wise convolutional neural network (CNN) to enhance the consideration of the relationship among the channels.
In short, Eq. 3 is expected to extract the foreground clean signal by maximizing the similarity between two inputs, where a trivial solution can be avoided \cite{Chen_CVPR2021}.
The one-side input $\ten{T'}$ is constrained by the low rank \cite{Bacca_AO2021}.
The other-side input $\ten{T}$ is not constrained by the rank.
This asymmetric operation boosts the extract of the common features \cite{Wang_CVPR2022}, i.e., the foreground clean signal, between the two inputs $\ten{T}$ and $\ten{T'}$.
In Eq. 4, each tensor and matrices are initialized with Tucker decomposition.

Moreover, to employ the channel-dependent sensitivity of DAS, the following objective function is used:
\begin{align}
  \vspace{0pt}
  \mathcal{L}_{\rm ChEnt}(\bm \beta, \bm \gamma) &= - \Biggl\{ \sum_{i=0}^{C-1} \frac{e^{\beta_{i}}}{\sum_{j=0}^{K-1} e^{\beta_{j}}} {\rm log} \Biggl( \frac{e^{\beta_{i}}}{\sum_{j=0}^{K-1} e^{\beta_{j}}} \Biggl)\nonumber\\
	& + \sum_{i=0}^{C-1} \frac{e^{\gamma_{i}}}{\sum_{j=0}^{K-1} e^{\gamma_{j}}} {\rm log} \Biggr( \frac{e^{\gamma_{i}}}{\sum_{j=0}^{K-1} e^{\gamma_{j}}} \Biggl) \Biggr\} / 2
  \label{eq:ch_entropy}
  \vspace{0pt}
\end{align}

\noindent
This function considers the difference in the noise level among the channels. 
The difference is implicitly enhanced by using the function.
We optimize the parameters of the proposed model using the following function, which is the combination of Eqs. 3 and \ref{eq:ch_entropy}:
\begin{align}
  \vspace{0pt}
  \mathcal{L} &= \mathcal{L}_{\rm F}(\ten{T}, \ten{Z}', \ten{T}', \ten{Z}) + \mathcal{L}_{\rm ChEnt}(\bm \beta, \bm \gamma)
	\label{eq:total_loss}
  \vspace{0pt}
\end{align}

\noindent
The model finally outputs $(\ten{T}+\ten{T}')/2$ as the denoised signal, which is obtained from iteratively updating the parameters with Eq. \ref{eq:total_loss}.
When $(\ten{T}+\ten{T}')/2$ includes the negative values, the elements are switched to zero values.
Note that our algorithm uses only noisy signals, i.e., not machine learning.

\vspace{0pt}
\section{Experiments}
\vspace{0pt}
\subsection{Experimental conditions}

\begin{figure}[t!]
  \centering
  \includegraphics[width=0.45\textwidth]{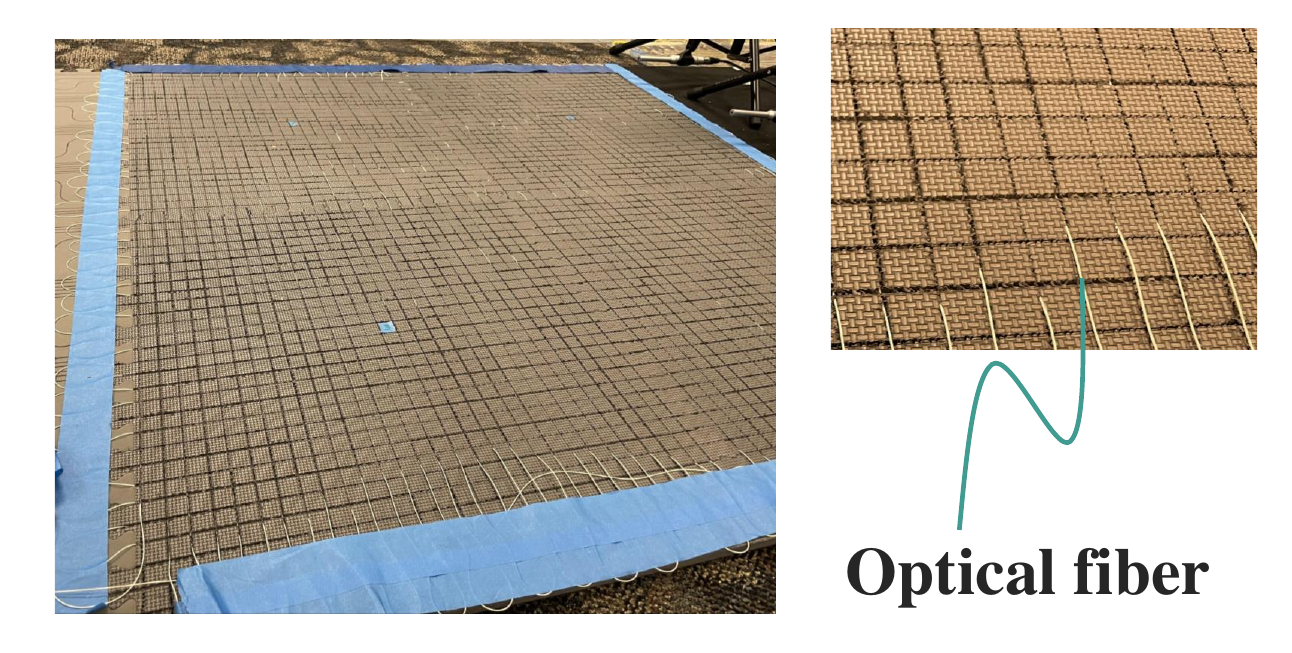}
  \vspace{-10pt}
  \caption{ Installation of optical fiber }
  \label{fig:carpet}
	\vspace{-5pt}
\end{figure}

\noindent
{\bf [Dataset, acoustic features, model parameters]}
To validate the proposed method, we conduct the acoustic experiment using DAS.
As shown in Fig. \ref{fig:carpet}, the optical fiber is embedded into a mat.
The size of the mat is 1.2m $\times$ 1.2m.
Sound source signals were played using a speaker and then omnidirectionally propagated to the optical fiber of the mat.
The straight-line distance between the edge of the mat and the speaker is about 2.8m.
We then obtained 50 channels from the DAS data of the mat.
For the sound sources, we used the ESC-50 dataset \cite{piczak_ACMM2015}, which is comprised of 2,000 audio clips.
The sampling frequency is 20kHz of DAS.
For the acoustic feature, we used the amplitude spectrogram for each channel, where the window or hop size is set to 640 or 320, respectively.
For the point-wise CNN of the ${\sf Predictor(\cdot)}$, the kernel size is set to 3 $\times$ 3.
In this work, the number of the iterations $N$ is set to 1500.
The parameters of the proposed method were updated by backpropagation using Adam \cite{Kingma_ICLR2015}, where the initial learning rate was set to $10^{-2}$.

\noindent
{\bf [Methods for comparison]}
For comparison, we used the following methods:

\begin{itemize}
	\item Spectral subtraction (SS) \vspace{-4.0pt}
	\item Singular value decomposition (SVD) \vspace{-4.0pt}
	\item Tucker decomposition \vspace{-4.0pt}
	\item Proposed method
\end{itemize}

\noindent
For SS, 5-minute ambient noise from DAS was used for calculating the noise floor.
The only spatial rank was restricted in SVD, Tucker decomposition, and the proposed method.
The frequency and time of the ranks were set to the full rank.
In SS, Tucker decomposition, and the proposed method, the denoised amplitude spectrogram was converted to its time domain using inverse STFT, where the phase of the noisy signal was used.

\noindent
{\bf [Evaluation metrics]}
For the evaluation, we conducted the 5-fold cross-validation using the ESC-50 dataset.
In the evaluation of denoising the DAS observation, the phase of the observed signal, i.e., the temporal lag among channels, for each channel is inaccessible.
For the consideration of the inaccessible phase, we define a cross-correlation improvement (CCi) between a played audio source signal and an observed DAS signal of a channel.
\begin{align}
  \vspace{0pt}
  {\rm CCi}_c := 20 \log_{10} \big(1+ {\rm CC}({\bf y}_{\rm denoised}^{(c)}, {\bf y}_{\rm source}) \big) \nonumber\\ / \big( 1+ {\rm CC}({\bf y}_{\rm noisy}^{(c)}, {\bf y}_{\rm source}) \big) 
	\label{eq:CCi}
  \vspace{0pt}
\end{align}

\noindent
where ${\rm CC}(\cdot, \cdot) \in [-1,1]$ is the maximum value of the cross-correlation (CC) between two signals.
${\bf y}_{\rm denoised}^{(c)}$ and ${\bf y}_{\rm noisy}^{(c)}$ are respectively denoised and noisy signals for channel $c$.
${\bf y}_{\rm source}$ is a played sound source signal.
Peak signal-to-noise ratio (PSNR) is also calculated for the evaluation of the proposed method.
\begin{align}
  \vspace{0pt}
  {\rm PSNR} := 20 \log_{10} \Biggl( {\rm max}({\bf y}) / \sqrt{ \frac{1}{T} \textstyle{\sum_{0}^{T-1}} \Big|  {\it y_t}^2 \Big| } \Biggr)
	\label{eq:PSNR}
  \vspace{0pt}
\end{align}

\noindent
where $\bf y \in \mathbb{R}^T$ represents a signal.
$y_t$ indicates the temporal index $t$ of $\bf y$.

\vspace{0pt}
\subsection{Experimental results}
\vspace{0pt}

\noindent
{\bf [Overall result of denoising]}
Table \ref{tbl:overall} shows the denoising performance for each method.
In this result, the predefined spatial rank of the low-rank-approximation methods was set to 1, respectively.
The result indicates that the proposed method outperforms the conventional methods in terms of each metric.
The denoising performance of conventional low-rank-approximation methods, i.e., SVD and Tucker decomposition, are particularly degraded in terms of CCi.
This is because rank$=$1 is a forceful prior, which leads to the limitation of the representational power of the model.
The proposed method is not significantly affected by such limitation since it is a DNN-based structure.

\begin{table}[]\
	\centering
	\caption{Overall result}
	\label{tbl:overall}
	\vspace{-5pt}
	\begin{tabular}{llcc}
	 &&& \\\hline
	 &&& \\[-8pt]
	 \multicolumn{2}{l}{}       & CCi {[}dB{]} & PSNR {[}dB{]}  \\[1pt]\hline
	&&& \\[-8pt]
	\multicolumn{2}{l}{SS}     & 0.002 & 16.26       \\ 
	&&& \\[-8pt]
	\multicolumn{2}{l}{SVD}  & -0.051 & 16.17      \\ 
	&&& \\[-8pt]
	\multicolumn{2}{l}{Tucker decomposition} & -0.013 & 17.57        \\
	&&& \\[-8pt]
	\multicolumn{2}{l}{Proposed}  & \bf 0.104 & \bf 19.56   \\\hline    
	\end{tabular}
	\vspace{5pt}
\end{table}

\begin{figure}[t!]
	\centering
	\includegraphics[width=0.49\textwidth]{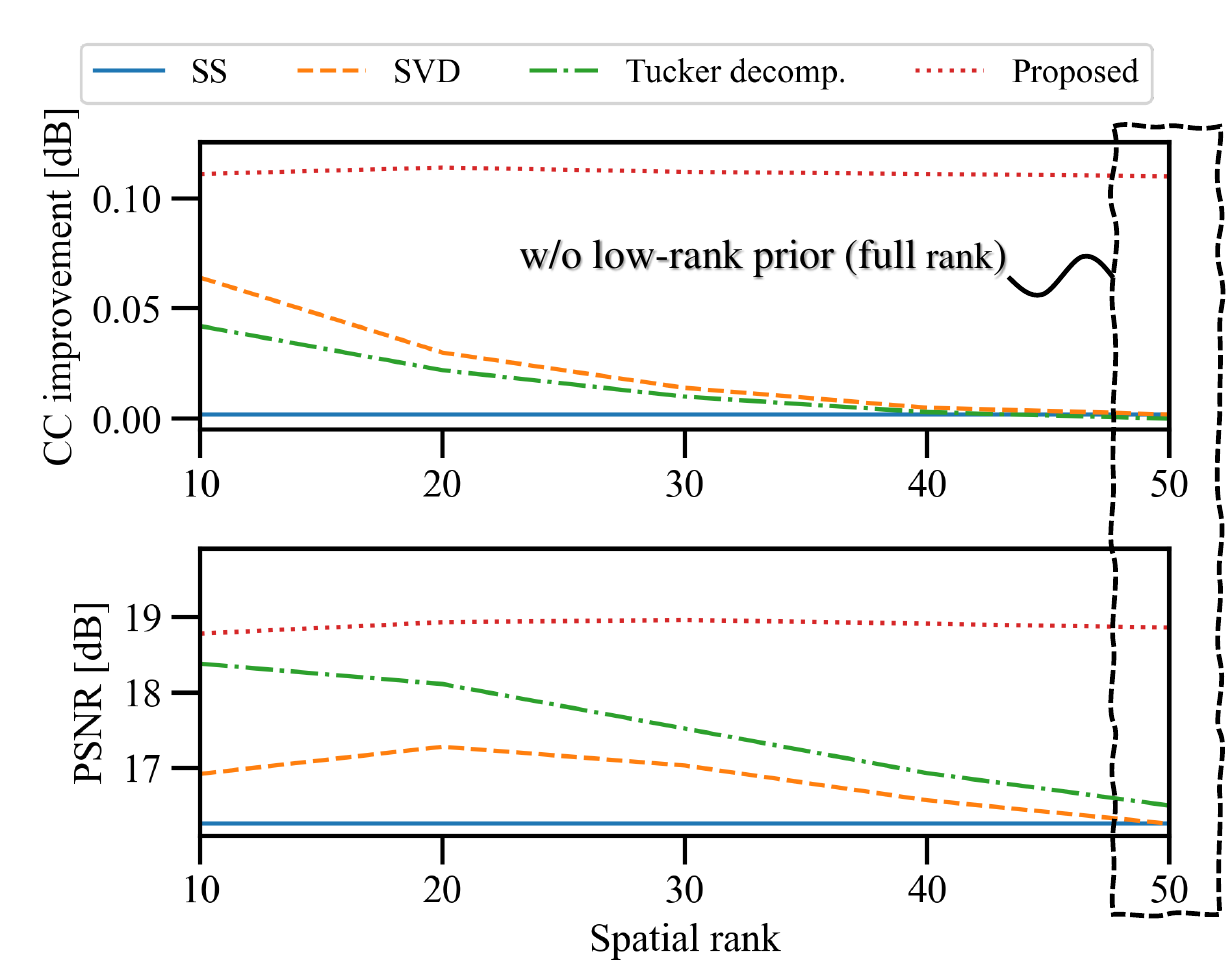}
	\vspace{-15pt}
	\caption{ Denoising performance with variable ranks }
	\label{fig:change_rank}
	\vspace{-5pt}
\end{figure}

\noindent
{\bf [Denoising performance with variable spatial ranks]}
In this section, we discuss the relationship between the denoising performance and the number of the spatial ranks.
In Fig. \ref{fig:change_rank}, the denoising performance with the variable spatial ranks is depicted.
The upper and lower figures respectively represent the performance in terms of CCi and PSNR.
Note that the performance of SS is not affected by the number of the spatial ranks. 
The result shows that the performance of the proposed method is robust to the change in the number of the spatial ranks compared with those of the conventional methods. 
When the number of the spatial ranks is larger, the performances of the conventional methods are decreased.
This is because the conventional methods access only the prior of the low rank in the spatial domain.
On the other hand, the proposed method utilizes the other prior, i.e., the channel-dependent sensitivity, in addition to the spatial low rank.
In particular, when the spatial rank is set to 50 (full rank), the proposed method still works as the great denoiser. 
The full rank in the spatial domain means that the methods cannot employ the prior of the spatial rank.
In other words, the prior of the channel-dependent sensitivity in Eq. \ref{eq:ch_entropy} is helpful for the denoising.  

\begin{figure}[t!]
  \centering
  \includegraphics[width=0.49\textwidth]{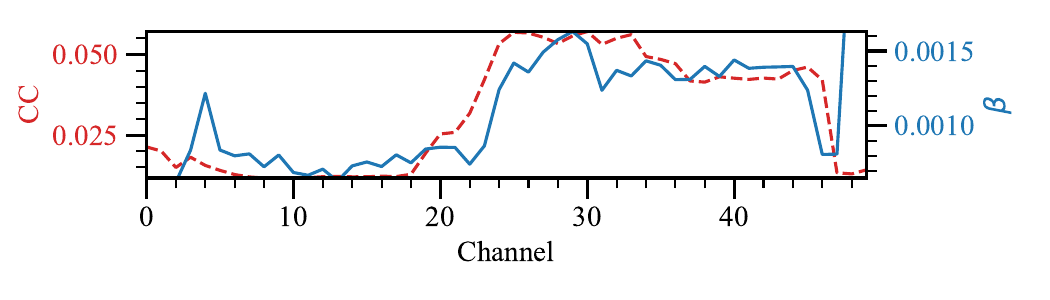}
  \vspace{-20pt}
  \caption{ Behavior of $\bm \beta$ and CC for interpretability }
  \label{fig:sensitivity_behavior}
	\vspace{-5pt}
\end{figure}

\noindent
{\bf [Difference in sensitivity among channels and interpretability]}
The dot-dash red line of Fig. \ref{fig:sensitivity_behavior} shows CC calculated with the noisy signals observed by DAS and the played sound source signals for each channel, where the average of all clips of the ESC-50 dataset is denoted.
In the blue line of Fig. \ref{fig:sensitivity_behavior}, the optimized $\bm \beta$ of the proposed method in Eq. \ref{eq:ch_subtraction} is shown.
As can be seen in the figures, the values of $\bm \beta$ where the sensitivities are worse in CC of Fig. \ref{fig:sensitivity_behavior} are larger compared with the other channels. 
This is because $\bm \beta$ works as a subtractor for extracting the foreground signal, i.e., the played audio.
In earlier channels with worse sensitivities, the noise level is higher than the observed audio signal.
In other words, there is no room for subtracting the noise.
The observed audio signal is zeroed out if the noise is subtracted from the noisy signal.
On the other hand, there is room for subtracting the noise in the later channels where the sensitivities are better compared with the other channels.
These result indicates that our proposed method contains the interpretability; a non-black box.
However, $\bm \beta$ of the around 50th channel takes unstable values.
The proposed method includes instability in some parameters because the convergence of DNNs for the global minimum is not guaranteed.

\begin{figure}[t!]
  \centering
  \includegraphics[width=0.49\textwidth]{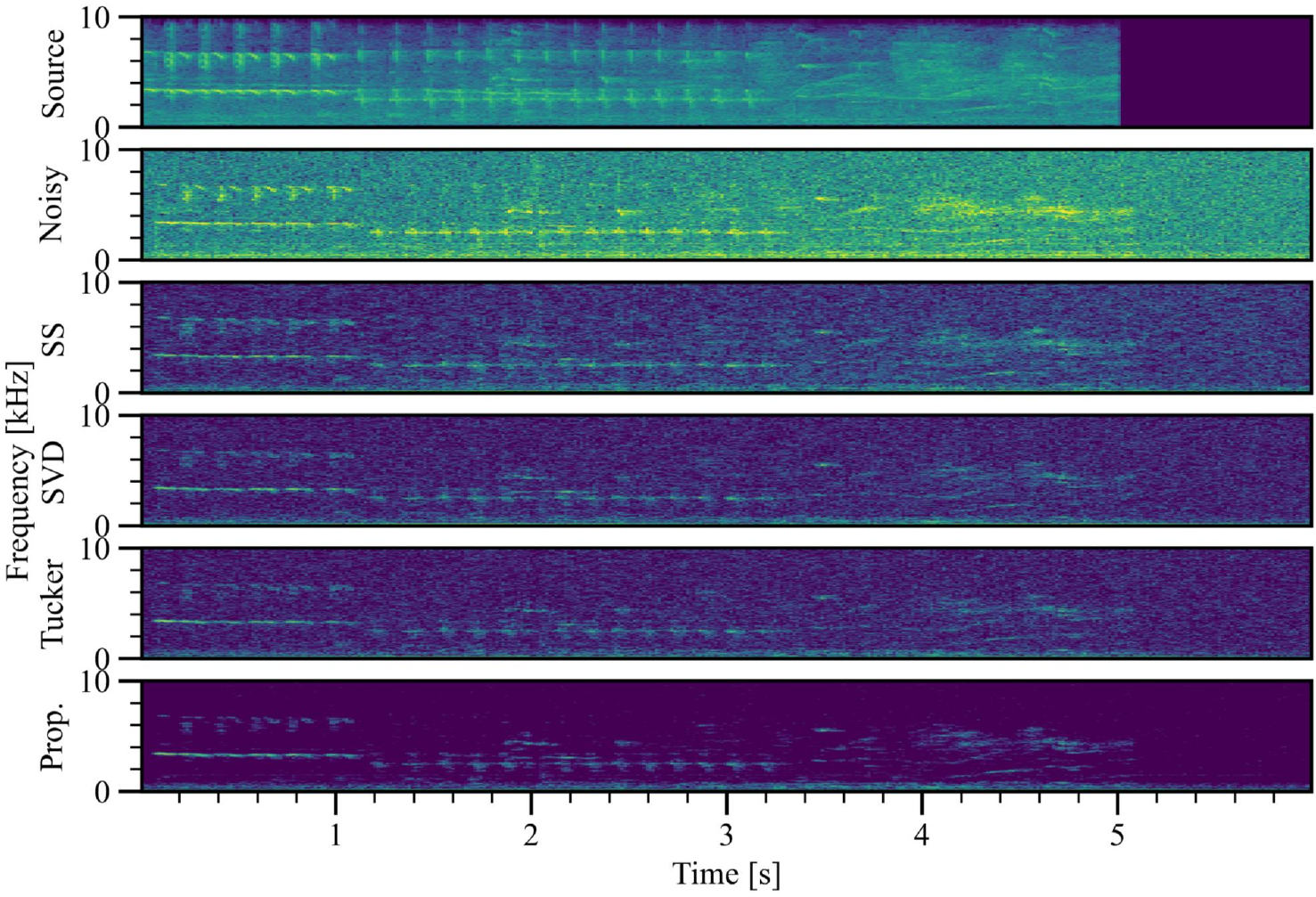}
  \vspace{-20pt}
  \caption{ Spectrograms of audio source, observed noisy signal, and denoised signals }
  \label{fig:spectrogram}
	\vspace{-5pt}
\end{figure}

\noindent
{\bf [Qualitative results of spectrograms]}
We finally confirm the qualitative results of the proposed method.
Fig. \ref{fig:spectrogram} shows the spectrograms of a selected sample, where the spatial rank is set to 10 and the channel index is 35.
Note that there is the temporal lag between the audio source and the other signals.
As can be seen in the figure, the noisy signal is heavily corrupted by the optical noise, that is, the Gaussian noise.
The proposed method removes the noise compared with the conventional methods.
However, the high frequency is distorted using the proposed method rather than the conventional methods. 
This is because $\beta_c$ in Eq. \ref{eq:ch_subtraction} depends on only the channel.
In other words, all of the frequencies are equally handled.
The lower frequency, which has more power than the high frequency, is prioritized in the proposed method.

\vspace{0pt}
\section{Conclusion}
\vspace{0pt}

In this paper, we proposed the DNN-structure signal-processing-based denoising method for DAS of optical fiber sensing.
The proposed method incorporates the flexibility of utilizing the priors and the interpretability.
The spatial low rank and the channel-dependent sensitivity are used as the priors of DAS.
The results show that the priors of the proposed method are helpful for denoising DAS data compared with the conventional methods.
In particular, the proposed method improved the denoising performance by 3.39dB in terms of PSNR compared with the conventional low-rank approximation.
Moreover, the model parameters of the proposed method reveal the relationship between the sensitivity and the parameters, the interpretability or non-black box.
Our proposed method will work in a more complicated environment with multiple sound sources.

\clearpage
\bibliographystyle{IEEEbib}
\bibliography{IEEEabrv,refs_et_al}

\begin{thebibliography}{10}

\bibitem{IP_OFC2021}
E.~Ip, Y.~Huang, M.~Huang, M.~Salemi, Y.~Li, T.~Wang, Y.~Aono, G.~Wellbrock,
  and T.~Xia,
\newblock ``Distributed fiber sensor network using telecom cables as sensing
  media: Applications,''
\newblock {\em Proc. Optical Fiber Communications Conference and Exhibition
  {\rm (}OFC{\rm )}}, pp. 1--3, 2021.

\bibitem{Ip_JOCN2022}
E.~Ip, J.~Fang, Y.~Li, Q.~Wang, M.~Huang, M.~Salemi, and Y.~Huang,
\newblock ``Distributed fiber sensor network using telecom cables as sensing
  media: technology advancements and applications,''
\newblock {\em Journal of Optical Communications and Networking {\rm (}JOCN{\rm
  )}}, vol. 14, no. 1, pp. 61--68, 2022.

\bibitem{Bouffaut_Frontiers2022}
L.~Bouffaut, K.~Taweesintananon, H.~Kriesell, R.~Rorstadbotnen, J.~Potter,
  M.~Landro, S.~Johansen, J.~Brenne, A.~Haukanes, O.~Schjelderup, and
  F.~Storvik,
\newblock ``Eavesdropping at the speed of light: Distributed acoustic sensing
  of baleen whales in the arctic,''
\newblock {\em Frontiers in Marine Science}, vol. 9, pp. 1--13, 2022.

\bibitem{Peter_JCSHM2021}
H.~Peter, X.~James, Z.~Shenghan, D.~Matthew, L.~Linqing, S.~Kenichi, P.~Carlo,
  Z.~Christian, M.~Demetrio, F.~Fabio, L.~Francisco, and M.~Chris,
\newblock ``Dynamic structural health monitoring of a model wind turbine tower
  using distributed acoustic sensing {(DAS)},''
\newblock {\em Journal of Civil Structural Health Monitoring}, vol. 11, pp.
  833--849, 2021.

\bibitem{Parker_FirstBreak2014}
T.~Parker, S.~Shatalin, and M.~Farhadiroushan,
\newblock ``Distributed acoustic sensing a new tool for seismic applications,''
\newblock {\em First Break}, vol. 32, no. 2, pp. 61--69, 2014.

\bibitem{Owen_EISIC2012}
A.~Owen, G.~Duckworth, and J.~Worsley,
\newblock ``Optasense: Fibre optic distributed acoustic sensing for border
  monitoring,''
\newblock {\em Proc. european Intelligence and Security Informatics
  Conference}, pp. 362--364, 2012.

\bibitem{Waagaard_OSAC2021}
O.~Waagaard, E.~R{\o}nnekleiv, A.~Haukanes, F.~Stabo-Eeg, D.~Thingb{\o},
  S.~Forbord, S.~Aasen, and J.~Brenne,
\newblock ``Real-time low noise distributed acoustic sensing in 171km low loss
  fiber,''
\newblock {\em OSA Continuum}, vol. 4, no. 2, pp. 688--701, 2021.

\bibitem{Lu_ICASSP2021}
Y.~Lu, Y.~Tian, S.~Han, E.~Cosatto, S.~Ozharar, and Y.~Ding,
\newblock ``Automatic fine-grained localization of utility pole landmarks on
  distributed acoustic sensing traces based on bilinear resnets,''
\newblock {\em Proc. {IEEE} International Conference on Acoustics, Speech and
  Signal Processing {\rm (}ICASSP{\rm )}}, pp. 4675--4679, 2021.

\bibitem{Bekara_GEOPHYSICS2007}
M.~Bekara and M.~Baan,
\newblock ``Local singular value decomposition for signal enhancement of
  seismic data,''
\newblock {\em GEOPHYSICS}, vol. 72, no. 2, pp. 59--65, 2007.

\bibitem{Chen_GEOPHYSICS2014}
Y.~Chen and J.~Ma,
\newblock ``Random noise attenuation by f-x empiricalmode decomposition
  predictive filtering,''
\newblock {\em GEOPHYSICS}, vol. 79, no. 3, pp. 81--91, 2014.

\bibitem{Zhao_GJI2021}
Y.~Zhao, Y.~Li, and N.~Wu,
\newblock ``Data augmentation and its application in distributed acoustic
  sensing data denoising,''
\newblock {\em Geophysical Journal International}, vol. 228, no. 1, pp.
  119--133, 2021.

\bibitem{Feng_TGARS2022}
Q.~Feng and Y.~Li,
\newblock ``Denoising deep learning network based on singular spectrum
  analysis―{DAS} seismic data denoising with multichannel svddcnn,''
\newblock {\em IEEE Transactions on Geoscience and Remote Sensing}, vol. 60,
  pp. 1--11, 2022.

\bibitem{Yang_GEOPHYSICS2023}
L.~Yang, S.~Fomel, S.~Wang, X.~Chen, W.~Chen, O.~Saad, and Y.~Chen,
\newblock ``Denoising of distributed acoustic sensing data using supervised
  deep learning,''
\newblock {\em GEOPHYSICS}, vol. 88, no. 1, pp. 91--104, 2023.

\bibitem{Zhou_JLT2013}
J.~Zhou, Z.~Pan, Q.~Ye, H.~Cai, R.~Qu, and Z.~Fang,
\newblock ``Characteristics and explanations of interference fading of a $\phi
  $ -{OTDR} with a multi-frequency source,''
\newblock {\em Journal of Lightwave Technology}, vol. 31, no. 17, pp.
  2947--2954, 2013.

\bibitem{Lempitsky_CVPR2018}
V.~Lempitsky, A.~Vedaldi, and D.~Ulyanov,
\newblock ``Deep image prior,''
\newblock {\em Proc. IEEE/CVF Conference on Computer Vision and Pattern
  Recognition {\rm (}CVPR{\rm )}}, pp. 9446--9454, 2018.

\bibitem{Zhong_PTL2019}
Z.~Zhong, F.~Wang, M.~Zong, Y.~Zhang, and X.~Zhang,
\newblock ``Dynamic measurement based on the linear characteristic of phase
  change in $\phi$-{OTDR},''
\newblock {\em {IEEE} Photonics Technology Letters}, vol. 31, no. 14, pp.
  1191--1194, 2019.

\bibitem{Tucker_psy1996}
L.~Tucker,
\newblock ``Some mathematical notes on three-mode factor analysis,''
\newblock {\em Psychometrika}, vol. 31, no. 3, pp. 279--311, 1966.

\bibitem{berouti_icassp1979}
Michael Berouti, Richard Schwartz, and John Makhoul,
\newblock ``Enhancement of speech corrupted by acoustic noise,''
\newblock {\em Proc. {IEEE} International Conference on Acoustics, Speech and
  Signal Processing {\rm (}ICASSP{\rm )}}, pp. 208--211, 1979.

\bibitem{Chen_CVPR2021}
X.~Chen and K.~He,
\newblock ``Exploring simple siamese representation learning,''
\newblock {\em Proc. IEEE/CVF Conference on Computer Vision and Pattern
  Recognition {\rm (}CVPR{\rm )}}, pp. 15750--15758, 2021.

\bibitem{Bacca_AO2021}
J.~Bacca, Y.~Fonseca, and H.~Arguello,
\newblock ``Compressive spectral image reconstruction using deep prior and
  low-rank tensor representation,''
\newblock {\em Applied Optics}, vol. 60, no. 14, pp. 4197--4207, 2021.

\bibitem{Wang_CVPR2022}
X.~Wang, H.~Fan, Y.~Tian, D.~Kihara, and X.~Chen,
\newblock ``On the importance of asymmetry for siamese representation
  learning,''
\newblock {\em Proc. IEEE/CVF Conference on Computer Vision and Pattern
  Recognition {\rm (}CVPR{\rm )}}, 2022.

\bibitem{piczak_ACMM2015}
J.~Piczak,
\newblock ``{ESC}: Dataset for environmental sound classification,''
\newblock {\em Proc. the 23rd {Annual ACM Conference} on {Multimedia} {\rm
  (}ACMM{\rm )}}, pp. 1015--1018, 2015.

\bibitem{Kingma_ICLR2015}
D.~Kingma and J.~Ba,
\newblock ``Adam: A method for stochastic optimization,''
\newblock {\em Proc. International Conference on Learning Representations {\rm
  (}ICLR{\rm )}}, 2015.

\end{thebibliography}
\end{document}